# Multi-beam phase mask optimization for holographic volumetric additive manufacturing

Chi Chung Li[a], Joseph Toombs[a], Vivek Subramanian[b,c], Hayden K. Taylor[a]

[a]Department of Mechanical Engineering, University of California, Berkeley, Berkeley, CA, USA 94720
[b]Department of Electrical Engineering and Computer Sciences, University of California, Berkeley, Berkeley, CA, USA 94720
[c]Institute of Electrical and Micro Engineering, Ecole Polytechnique Federale de Lausanne, Lausanne, Vaud, Switzerland

## ABSTRACT

The capability of holography to project three-dimensional (3D) images and correct for aberrations offers much potential to enhance optical control in light-based 3D printing. Notably, multi-beam multi-wavelength holographic systems represent an important development direction for advanced volumetric additive manufacturing (VAM). Nonetheless, searching for the optimal 3D holographic projection is a challenging ill-posed problem due to the physical constraints involved. This work introduces an optimization framework to search for the optimal set of projection parameters, namely phase modulation values and amplitudes, for multi-beam holographic lithography. The proposed framework is more general than classical phase retrieval algorithms in the sense that it can simultaneously optimize multiple holographic beams and model the coupled non-linear material response created by co-illumination of the holograms. The framework incorporates efficient methods to evaluate holographic light fields, resample quantities across coordinate grids, and compute the coupled exposure effect. The efficacy of this optimization method is tested for a variety of setup configurations that involve multi-wavelength illumination, two-photon absorption, and time-multiplexed scanning beam. A special test case of holo-tomographic patterning optimized 64 holograms simultaneously and achieved the lowest error among all demonstrations. This variant of tomographic VAM shows promises for achieving high-contrast microscale fabrication. All testing results indicate that a fully coupled optimization offers superior solutions relative to a decoupled optimization approach.

## 1 INTRODUCTION

Volumetric additive manufacturing (VAM) is emerging as an attractive 3D printing method[1,2]. In contrast to point-scanning or plane-scanning methods, VAM effectively patterns a unit volume at a time (and therefore more precisely referred to as volume-at-once methods). This mode of patterning offers numerous fabrication advantages such as fast printing speed, ability to overprint onto existing structures[3], compatibility with a wide range of materials[4], and low surface roughness (down to $R_a$ of 6 $nm$[5]) for fabricating optical components[5,6].

VAM methods such as orthogonal projection[2] and tomographic reconstruction[1] mainly operate in a low numerical aperture (NA) regime where beam convergence or divergence can be minimized to negligible levels over the patterning volume. When scaling these techniques down to micron or sub-micron scales where high NA imaging is required, the diffraction effect starts to become significant, and the depth of focus (DOF) becomes much smaller than the patterning volume. Optical blurring in the defocused regions substantially reduces the geometric fidelity of the print. To estimate the severity of this problem, one can use the Abbe axial resolution as a rough measure of the DOF of an image and the Abbe lateral resolution as a length reference of the patterning volume. As shown in Figure 1, the DOF (axial resolution) is shrinking rapidly with numerical aperture. It reduces from about 50 times of the lateral resolution at 0.1 NA to only 6 times of lateral resolution at 0.7 NA. Since the patterning volume typically has a lateral size that is hundreds of times that of lateral resolution of the beam, it means that a majority of the patterning region is out of the DOF at 0.7 NA. To remedy this problem, these patterning systems can leverage holographic projection to axially focus light onto all object points and keep them in focus. As an added benefit, these systems can use the same phase modulation device that creates the holograms to also compensate for aberrations in the optical train[7]. Some aberrations, such as those introduced by refractive index interfaces, become more prominent as NA increases[8,9]. From these perspectives, holography is a crucial tool to help scale down the above two types of VAM systems.

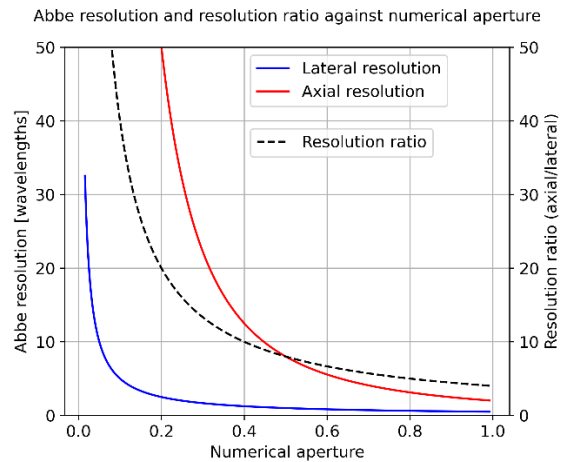

Figure 1 Lateral and axial Abbe resolution limit, and resolution ratio

On the other end, there exist VAM systems[10–13] that purely rely on the 3D focusing power of holographic projection to introduce spatial selectivity. Instead of using a multi-beam or a scanning beam configuration, these holographic VAM systems only pattern with a single stationary holographic beam. Due to the optical resolution anisotropy shown in Figure 1, the resolution and feature size limit of the printed object are also bound to be anisotropic. This anisotropy limits the geometrical freedom of printable objects[11,12] and this problem is much more severe at low NA[13]. For these systems, transitioning to a multi-beam configuration would reduce directional dependence of resolution and improve overall reconstruction fidelity. In particular, transitioning to multi-wavelength systems opens up special resolution enhancement possibilities, such as two-color photo-initiation/inhibition[14–16], STED[17–19] (stimulated emission depletion), or two-color two-step absorption[20]. Currently, these multi-wavelength photoexcitation mechanisms are not yet tested on holographic patterning setups. Overall, multi-beam holographic VAM is a relatively less explored area with much potential to achieve higher speed and resolution than current single-beam holographic systems.

Multi-beam holographic VAM is an attractive hybrid of the above techniques which leverage both the focal point steering capability of holography and the concept of synthetic aperture in multi-beam approaches. In such systems, the user needs to design the holographic projections that recreate the desired object within the patterning material. Since holograms are typically created by applying a spatial phase modulation to the beam, the above design problem is the problem of designing phase modulation masks. This inverse design problem in the field of computer-generated holography (CGH) is ill-posed and non-trivial. Not all target light distribution is physically realizable because light propagation needs to follow the law of electromagnetism. In addition, the degrees of freedom in the target light distribution often far exceeds that in the design parameters, and therefore there are no exact nor unique solutions in most situations. The class of algorithms used to solve this problem in holographic imaging is called phase retrieval algorithms and one of the most prominent examples is the Gerchberg-Saxton algorithm[21,22]. Although there are a variety of phase retrieval methods[23–26], so far they only consider single-beam systems and do not model material responses.

In VAM, the typical response of interest is the local degree of conversion (DOC). In general, DOC exhibits a non-linear relationship with exposure dose and this relationship needs to be properly accounted for during optimization. The material response relationship is particularly complex when multi-photon or multi-wavelength photoexcitation mechanisms are involved. For example, the amount of photoexcitation in multi-photon holographic lithography[10–12,27–30] has a quadratic or higher-order dependence on illumination intensities. Further, the aforementioned photoinhibition system[14–16] and STED lithography[17–19] introduce an antagonistic mechanism between the illumination of different colors and therefore impose a subtractive relationship between the exposure effects. Alternately, there could be a synergistic relationship between exposures in two-color two-step absorption[20] or with spiropyran photoswitch[31]. In general, systems that leverage these mechanisms require a computation layer in the phase mask design process to account for non-linearities and couple different exposure effects together. Unfortunately, such layer is missing from existing phase retrieval methods.

This work proposes a non-convex optimization method to co-optimize the spatial phase modulation masks for multiple holographic beams such that the overall material response matches a specified target as closely as possible. In this proposed optimization framework, a flexible photoexcitation model is used to approximate the aggregate effect of multi-wavelength exposure and multi-photon absorptions. Additionally, this framework includes a differentiable resampling step that allows quantities to be evaluated at distinct sampling rates and facilitates efficient gradient-based optimization. Apart from static projections, the proposed optimization framework also can optimize mask time-multiplexed holographic beams that are stationary or moving relative to the material. Therefore, the applicability of this framework covers dynamic hologram playback for speckle reduction and holo-tomographic patterning configurations.

Section 2 introduces the major components of the multi-beam optimization framework. Section 3 presents the results from various test cases and discusses the findings.

## 2 MULTI-BEAM OPTIMIZATION FRAMEWORK

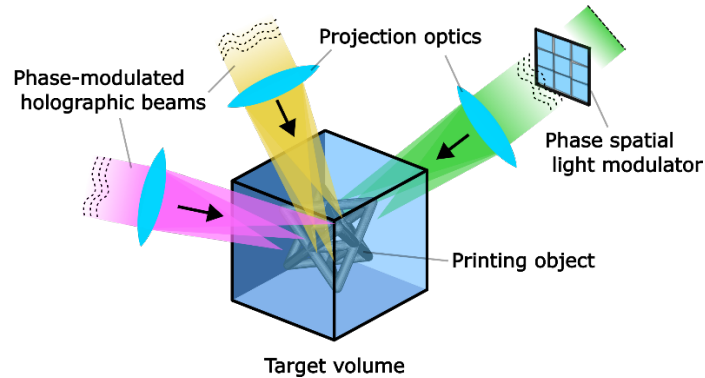

Figure 2 Conceptual illustration of multi-beam holographic volumetric additive manufacturing

Figure 2 is a concept drawing of multi-beam multi-color holographic VAM. In the printing process, illumination from multiple angles and at different wavelengths delivers a coupled photoexcitation to the material. The photoexcitation triggers photopolymerization and locally increases the degree of conversion (DOC) of the material. Generally, the desired response can be a prescribed DOC or any of the dependent properties such as elastic modulus and refractive index. The optimization aims to minimize the difference between the response target $f_T$ and the material response $f_m = \mathcal{M}(f)$ resulted from photoexcitation $f$. The overall reconstruction error is quantified by a loss function and reported as a scalar loss value.

Besides the loss function, other components of this framework include a coherent propagation model, a resampling operation, a photoexcitation model, and a material response model. In this multi-beam optimization, there are practical challenges in modeling and computing the coupling effect of each of the holographic beams in a numerically consistent manner. The differentiable resampling step and the polynomial photoexcitation model are used to address these problems that uniquely pertain to multi-beam configurations.

Figure 3 outlines the algorithmic structure of the framework. It also depicts how information flows through various components of the optimization framework in the forward simulation (blue arrows) and in the computation of loss gradient with automatic differentiation (orange arrows).

The definitions of the components and variables on the flowchart are described in the following subsections. Section 2.1 introduces the band constraint loss function and the iterative solution update scheme. Section 2.2 details the coherent propagation method that simulates holographic light fields based on the input variables such as phase modulation values. Section 2.3 describes the resampling step that maps quantities across computational grids and section 2.4 discusses the automatic differentiability of the operations that enable efficient gradient evaluations. Section 2.5 discusses how general inter-beam coupling at the material response level can be represented in the proposed photoexcitation model.

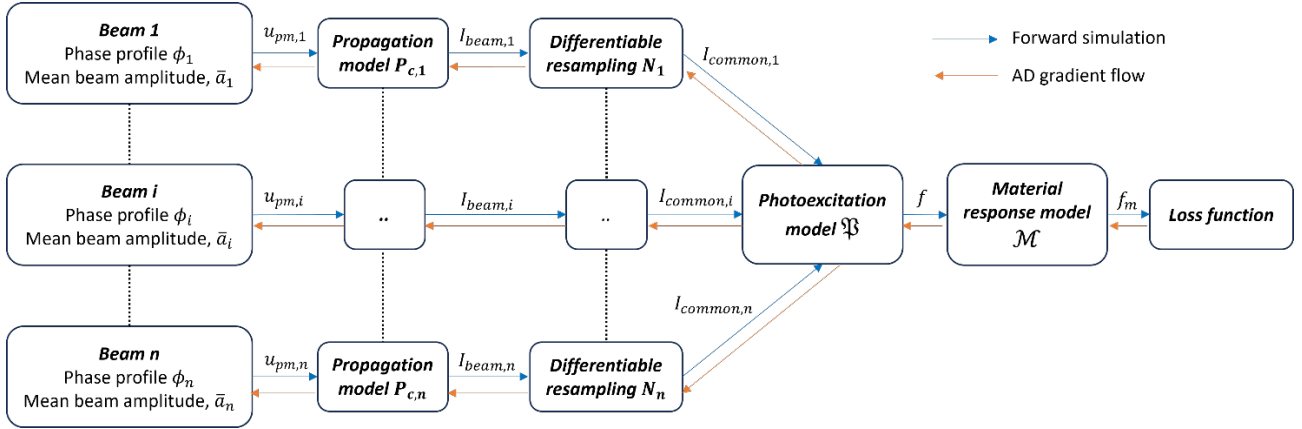

Figure 3 Flow chart showing the components in the proposed framework and direction of information flow in the forward simulation (blue arrows) and direction of gradient flow in automatic differentiation (AD) (orange arrows).

## 2.1 Loss function and gradient descent update

This optimization framework formulates all optimization goals directly into a loss function. Recently, a band constraint $L_p$-norm (BCLP) loss function[32] has been found to be able to flexibly prescribe accuracy requirements and priorities in the design of incoherent optical projection for tomographic VAM. This framework applies the BCLP loss function to the current context with multi-beam holographic reconstruction. The general form of the BCLP loss function $\mathcal{L}$ is expressed as:

$$\mathcal{L} = \left( \int_V w(\underline{r}) \left| |f_m(\underline{r}) - f_T(\underline{r})| - \varepsilon(\underline{r}) \right|^p d\underline{r} \right)^{\frac{q}{p}}, \quad (1)$$

with spatial position $\underline{r} = [x, y, z]$, optimization weights $w$, error tolerance $\varepsilon$, $L_p$ norm parameter $p$, and exponent $q$ defined in the original text[32]. The integral limit $V = \{\underline{r} : |f_m(\underline{r}) - f_T(\underline{r})| > \varepsilon(\underline{r})\}$ is set to only cover regions where the error $|f_m(\underline{r}) - f_T(\underline{r})|$ exceeds tolerance $\varepsilon(\underline{r})$. As discussed in original text[32], the weight and tolerance parameters are used to specify local accuracy requirements and obtain solutions that better align with application goals. To facilitate an intuitive analysis of optimization results, all demonstrations presented in this work use a special case of the above general BCLP loss function that measures the squared error:

$$\mathcal{L} = \int_\infty \left( f_m(\underline{r}) - f_T(\underline{r}) \right)^2 d\underline{r}, \quad (2)$$

by setting $w = 1$, $\varepsilon = 0$, and $p = q = 2$. In other words, all regions in space have equal weighting and zero error tolerance.

The optimization variables for beam $i$ are the phase modulation profile $\phi_i$ on the 2D phase masks and the mean beam amplitude $\bar{a}_i$. At a high level of abstraction, the optimization problem read as:

$$\min_{\substack{\underline{\phi} \\ \underline{\bar{a}}}} \mathcal{L} = \int_\infty \left( f_m(\underline{r}) - f_T(\underline{r}) \right)^2 d\underline{r}, \quad (3)$$

where $\underline{\phi}$ is the vector containing phase modulation profiles for all beams and $\underline{\bar{a}}$ is the vector containing mean beam amplitudes for all beams.

As depicted in Figure 3, evaluating the material response $f_m$ involves multiple computation steps. The propagation model $P_{c,i}$ first computes the resultant 3D holographic light field $u_{h,i}$ from the input parameters $\phi_i$ and $\bar{a}_i$. The intensity of the hologram is taken as the squared modulus of the complex light field and is expressed as $I_{beam,i} = |u_{h,i}|^2$. A differentiable interpolation step samples the intensity values $I_{beam,i}$ (evaluated at beam-dependent coordinate grids) on a common coordinate grid and yield $I_{common,i}$. The polynomial photoexcitation model takes all intensity fields $I_{common,i}$ as arguments and evaluates the effective photoexcitation $f$. Finally, the material response model $\mathcal{M}$ maps the photoexcitation to material response values $f_m = \mathcal{M}(f)$.

The optimization starts with an initial guess of the solution and then iteratively updates the solution with a gradient descent stepping method. In each iteration (indexed by $k$), the optimization program computes the loss value $\mathcal{L}_k$ and the loss gradient $\nabla \mathcal{L}_k$ with respect to all optimization variables. The algorithm updates the trial solution for the next iteration by:

$$\underline{\boldsymbol{\phi}}_{k+1} = \underline{\boldsymbol{\phi}}_k - \eta_{\underline{\phi}} \nabla_{\underline{\phi}} \mathcal{L}_k \tag{4}$$

and

$$\underline{\overline{\boldsymbol{a}}}_{k+1} = \underline{\overline{\boldsymbol{a}}}_k - \eta_{\underline{\bar{a}}} \nabla_{\underline{\bar{a}}} \mathcal{L}_k, \tag{5}$$

where $\nabla_{\underline{\phi}}$ and $\nabla_{\underline{\bar{a}}}$ are gradient with respect to $\underline{\boldsymbol{\phi}}$ and $\underline{\overline{\boldsymbol{a}}}$ respectively, and $\eta_{\underline{\phi}}$ and $\eta_{\underline{\bar{a}}}$ are the step size for $\underline{\boldsymbol{\phi}}$ and $\underline{\overline{\boldsymbol{a}}}$ respectively. In other words, all phase modulation values use the same step size and all mean beam amplitudes use the same step size. All step sizes are chosen such that the algorithm updates the initial solution by approximately 0.1% of its original mean magnitude upon the completion of the first iteration. Although the loss values typically converge in less than 2000 iterations under this step size, all optimizations are performed for 5000 iterations for fair comparisons of their asymptotic behavior.

## 2.2 Coherent light propagation with Angular Spectrum Method

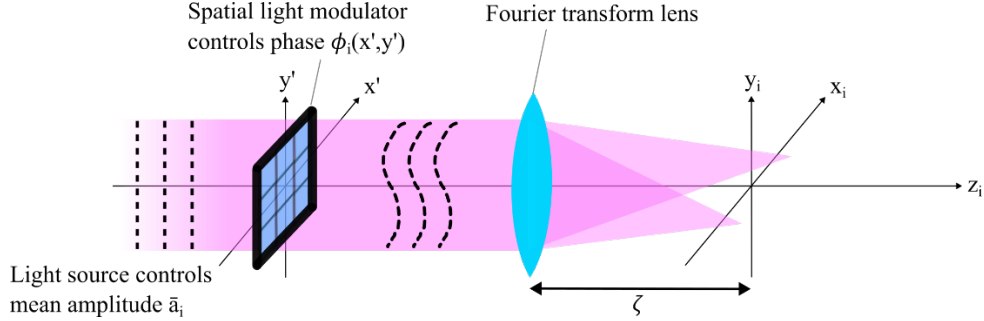

Figure 4 Projection of Fourier hologram and associated coordinate axes

During the optimization, the propagation model is used to predict the 3D intensity field of the holographic projection based on the complex amplitude of the light field at the phase modulation plane. The work focuses on optimizing phase-modulated Fourier holograms. As illustrated in Figure 4, Fourier holograms are projected near the back focal plane of the Fourier transform lens. Fresnel holograms can be similarly modelled and optimized with the proposed framework, but all demonstrations in this paper use a mathematical expression that computes light fields for Fourier holograms.

The propagation model assumes each of the beams is monochromatic and maintains perfect coherence within the beam. Since the illumination of each of the beams in the system can have a different wavelength and potentially be separated from each other in time, there may not be inter-beam interference effects. This work takes the limit where the beams are effectively mutually incoherent so that the beams can only be coupled at the photoexcitation model in terms of intensity (described in section 2.5). As mentioned in section 2.3, this assumption greatly simplifies the sampling requirements in the interpolation step. This section details the method used in simulating the projection of one holographic beam (beam $i$).

In this model, the phase modulation profile is denoted by $\phi_i(x', y')$ and the beam amplitude is denoted by $a_i(x', y')$, where $x'$ and $y'$ are in-plane spatial coordinates at the phase modulation plane. $\phi_i(x', y')$ is assumed to have a range of $[0, 2\pi)$. $a_i(x', y')$ is assumed to carry a constant mean-normalized amplitude profile $\hat{a}_i(x', y')$ with mean beam amplitude $\bar{a}_i$ being adjustable as an optimization variable. Therefore, the light field amplitude leaving the phase modulation plane can be expressed as:

$$u_{pm,i}(x', y') = \bar{a}_i \hat{a}_i(x', y') e^{j\phi_i(x', y')}, \tag{6}$$

with $j$ being the imaginary unit. For simplicity, the examples in this paper assume a flat-top beam profile where $\hat{a}_i(x', y') = 1$.

The coherent propagation model is constructed with Angular Spectrum Method (ASM) in Fourier optics[33]. A Fourier transform relates the light fields at one focal plane of the Fourier transform lens to the other. The ASM relates the light field at a chosen plane to the light field at other parallel planes of interest.

After simplification, the holographic light field at the target planes $u_{h,i}$ can be expressed as:

$$u_{h,i}(x_i, y_i, z_i) = \frac{e^{jk2\zeta}}{j|\lambda\zeta|} \iint_{-\infty}^{\infty} u_{pm,i}(x', y') H(x', y', z_i) e^{-j2\pi(f_x x' + f_y y')} dx' dy' \bigg|_{\substack{f_x = \frac{x_i}{\lambda\zeta} \\ f_y = \frac{y_i}{\lambda\zeta}}} \quad (7)$$

$$= \frac{e^{jk2\zeta}}{j|\lambda\zeta|} \mathcal{F} H(x', y', z_i) u_{pm,i}(x', y') \bigg|_{\substack{f_x = \frac{x_i}{\lambda\zeta} \\ f_y = \frac{y_i}{\lambda\zeta}}} = P_{c,i} u_{pm,i}, \quad (8)$$

where $\lambda$ is the wavelength of the light, $k = \frac{2\pi}{\lambda}$ is the wavenumber, $\zeta$ is the effective focal length of the Fourier Transform lens, $H(x', y', z_i)$ is the overall transfer function relating the light field at the phase mask plane to the hologram plane of interest $z$, $\mathcal{F}$ is Fourier transform operator and $P_{c,i}$ is the overall coherent propagator that is linear in complex amplitude $u_{pm,i}$. The holographic light field $u_{h,i}$ is parametrized by the local coordinate system of the beam $[x_i, y_i, z_i]$, where $x_i \parallel x'$, $y_i \parallel y'$, and $z_i$ remains parallel to the optical axis. On equation (7) and (8), $\lambda, k, \zeta$ and $H$ can take different values in different beams and their subscript $i$ are hidden for readability. For computational convenience, the constant $\frac{e^{jk2\zeta}}{j|\lambda\zeta|}$ is dropped to reduce the order of magnitudes in floating point arithmetics and avoid overflow. As mentioned earlier, the intensity $I_{beam,i}$ of the beam is taken as $I_{beam,i} = |u_{h,i}|^2$.

The complex transfer function depends on the reconstruction depth $z$ and the optical properties of the propagation medium. When propagating the beam through multiple materials with different refractive indices, refractions at the index interfaces would introduce axial shifts and aberrations on the focal point. Furthermore, if the material is attenuative, different spatial frequency components of the holographic beam would also suffer different energy losses due to their different travel distances. An accurate propagation model would need to account for these non-idealities.

In this work, all material interfaces are assumed to be perpendicular to the optical axis and intersecting the entire beam. The overall transfer function $H$ is composed of $s$ incremental propagations $\Delta H_m$ with each $\Delta H_m$ modeling the propagation in a specific medium $m$. The overall transfer function is written as:

$$H(x', y', z) = \prod_m^s \Delta H_m(x', y', \Delta z_m), \quad (9)$$

where $\Delta z_m$ is the incremental propagation distance in medium $m$ along the optical axis. The incremental propagations start from the free-space back focal plane of the Fourier Transform lens, to each of the relevant index interfaces, and finally end at the plane of interest $z$. Using the transfer function in ASM[33] and including an attenuation term, the incremental transfer function $\Delta H_m$ is written as:

$$\Delta H_m(x', y', \Delta z_m) = \exp\left(-\frac{\bar{\alpha}_m \Delta z_m}{2\gamma(x', y')} + j k_m \Delta z_m \gamma(x', y')\right), \quad (10)$$

where $\gamma(x', y') = \sqrt{1 - \left(\frac{x'}{n_m \zeta}\right)^2 - \left(\frac{y'}{n_m \zeta}\right)^2}$ is the directional cosine of the angular spectrum component relative to the optical axis, $n_m$ is the refractive index of material $m$, and $\bar{\alpha}_m$ is the attenuation coefficient of material $m$. The value of $\Delta H_m$ is set at zero at the far $x', y'$ regions where $\gamma$ becomes imaginary because they correspond to non-propagating (evanescent waves) in ASM[33]. All example holograms in this work are projected from air, through a 1.25 mm fused silica glass (with an index of 1.464 and negligible attenuation), into a polymer precursor material with an index of 1.516 and attenuation coefficient of 0.2 $mm^{-1}$. The center of the hologram is positioned 1.5mm deep into the precursor material.

On the phase modulation plane, $u_{pm,i}$ is sampled at the same sampling rate and extent as the spatial light modulator at 1 sample/pixel. By the reciprocal relationship of the Fourier transform, the holographic light field $u_{h,i}(x_i, y_i, z_i)$ is naturally sampled at the diffraction-limited resolution along lateral dimensions ($x_i$ and $y_i$) with lateral extent matching the size of the individually addressable volume. The sampling rate of $u_{h,i}$ along $z_i$ is user-defined. For efficient sampling, the axial sampling rate is chosen to be the same as the axial diffraction-limited resolution in the medium. Therefore, the sampling rates inherit the same anisotropy as the Abbe diffraction limits in the beam's local coordinate system.

## 2.3 Resampling via interpolation

In the optimization framework, each contributing beam can have distinct configurational parameters such as wavelength, phase modulation pitch, phase modulation resolution, beam position, and beam orientation relative to the global coordinate system. All these parameters influence the spatial sampling rate and sampling location of the simulated intensity of the holographic light field $I_{beam,i}$. Generally, the volumetric sampling grid of the hologram projected by each individual beam would not match the grids of other holograms. Therefore, a resampling step is necessary to approximate all holographic intensity fields at a common coordinate grid where their coupled contributions and the loss function can be evaluated.

In the demonstrations, a linear interpolation step resamples the holographic intensity fields from their individual beam-dependent coordinate grids onto a common coordinate grid. The resampling operation for the $i$-th beam's intensity $I_{beam,i}$ is represented by:

$$I_{common,i}(\mathbf{r}) = N_i \left( I_{beam,i}(x_i, y_i, z_i) \right), \tag{11}$$

where $I_{common,i}$ is the resampled intensity value and $\mathbf{r} = [x, y, z]$ is the position vector in the common coordinate system.

An additional benefit of performing such resampling is that the spatial sampling rates in either coordinate system can be individually controlled for the best use of computing resources. For example, the holographic intensity fields can use a diffraction-limited anisotropic sampling rate while the response target $f_T$ and its reconstruction error can use an application-driven (and potentially isotropic) sampling rate.

The resampling is purposely done after the amplitude and phase information of the light field are separated. Otherwise, resampling the complex light field $u_{h,i}$ in real and imaginary parts (where amplitude and phase are mixed) will cause aliasing in phase and cancellation problems. The spatial period at which the phase fluctuates can be as short as the wavelength of the light (often in hundreds of $nm$) which is much shorter than the axial diffraction limit (often in $\mu m$ or tens of $\mu m$ unless $NA \to 1$ and $n_m \to 1$). Therefore, accurate resampling of the complex light field would often require the field itself to be discretized at an impractical sampling rate. Even with a resampling rate slightly above the Nyquist limit, cancellations would happen when linearly interpolating real and imaginary parts separately. In view of this, the current optimization framework limits coherent interaction to happen only within the propagation model of each of the beams.

## 2.4 Auto-differentiation for loss gradient evaluation

As in high-dimensional inverse design problems, the use of numerical differentiation such as finite difference method to evaluate loss gradient $\nabla \mathcal{L}$ is not computationally tractable in this context due to the sheer number (in order of millions) of forward simulations required. On the other extreme, symbolic differentiation provides an explicit expression of the loss gradient with the evaluation cost often being similar to the forward simulation itself. In theory, one can derive such an expression by applying the chain rule of differentiation and multiplying the gradients of each of the operations symbolically. Most of the mathematical operations in this framework (such as the loss function, material response model, photoexcitation model, and coherent propagation model) have gradients of simple forms. However, the resampling step mentioned in section 2.3 is one of the example operations where manual symbolic differentiation and custom implementation are tedious and likely not performant in high-level programming languages.

For this reason, this work performs all computations (including the resampling step) with an automatic differentiation (AD) library PyTorch. At its core, the AD library provides a set of elementary operations (such as arithmetic and trigonometric functions) that facilitate the numerical evaluation of analytical gradient of function outputs with respect to all function inputs. By combining these differentiable elementary operations, the library also facilitates the differentiation of high-level operations (such as linear interpolation) and custom functions (such as the loss function). The gradient of these operations can be found by programmatically executing the chain rule of differentiation. In this work, the loss gradients with respect to the optimization variables ($\nabla_{\underline{\phi}}\mathcal{L}$ and $\nabla_{\underline{a}}\mathcal{L}$) are directly computed by AD in PyTorch's backend. As a result, both the forward simulation and the gradient evaluation in this study can be computed on the GPU efficiently.

## 2.5 Material response model and polynomial effective photoexcitation

As mentioned in the introduction, the relationship between material response and light exposure can be complicated. Firstly, the change in degree-of-conversion of the polymer and hence other dependent properties are not directly proportional to the amount of photoexcitation. The presence of induction period, auto-acceleration effects, and saturation naturally creates non-linearities in the process. Furthermore, the use of super-resolution schemes often involves novel mechanisms where the amount of photoexcitation varies with illumination intensity in a superlinear or wavelength-dependent manner. To model this complicated relationship between illumination and material response in optimization, this work proposes a simple composite model to capture these important features concisely and flexibly.

The proposed composite model consists of two layers, namely a material response model $\mathcal{M}$ and a photoexcitation model $\mathfrak{P}$. The response $f_m$ is expressed as:

$$f_m = \mathcal{M}(\mathfrak{P}(\underline{I})) = \mathcal{M}(f), \tag{12}$$

where $\underline{I}$ is the intensity vector with each element being the intensity of each holographic beam $I_{common,i}$, and the intermediate variable $f = \mathfrak{P}(\underline{I})$ represents the effective photoexcitation.

At the upper level, the material response model $\mathcal{M}$ relates the amount of photoexcitation $f$ that the material receives to the degree of the chosen response $f_m$ (such as the degree of conversion or elastic modulus). $\mathcal{M}$ is a deterministic univariate function of photoexcitation $f$ and it is primarily responsible to model the non-linearity in polymerization kinetics. This definition of the response model is consistent with its corresponding model in a tomographic projection optimization framework[32] previously proposed by the authors of this paper. As discussed in previous work, the mathematical expression of $\mathcal{M}$ should preferably be chosen to be differentiable and invertible for optimization purposes. For instance, the logistic function, affine function, and identity function all satisfy these properties.

At the lower level, the photoexcitation model $\mathfrak{P}$ relates the intensities (or irradiance) $I_{common,i}$ of all holographic beams to the effective photoexcitation $f$ delivered to the material. The proposed framework expresses the effective photoexcitation as a multivariate polynomial of the intensities for several reasons. Firstly, the polynomial form can directly express common theoretical and empirical relationships developed in various super-resolution schemes. For example, the theoretical probability of occurrence of two-photon absorption depends on illumination intensity (photon flux) quadratically[34,35,27,36,37]. In two-color photoinitiation and photoinhibition systems[14,15,38] where blue light initiates the polymerization reaction and ultraviolet light inhibits the reaction, the overall polymerization rate is also found to be a function of the difference of the two intensities. In fact, relationships with more sophisticated expressions can also be readily approximated by polynomials through Taylor series expansion. From this point of view, multivariate polynomial is a natural and relevant choice of expression in the current context. Secondly, polynomials are practical because they are concise, highly interpretable, and differentiable. Empirically, conciseness reduces the number of parameters to be fitted and facilitates mechanistic understanding. Since the effect of each of the coefficients in the polynomial is well understood mathematically, users can also limit the order of the polynomial based on known information. As in other components of the framework, smoothness and differentiability facilitate efficient gradient-based optimization.

Demonstrations in this work limit the polynomial order to be 2 or below and write $f$ as:

$$f = \mathfrak{P}(\underline{I}) = \underline{c_1}^T \underline{I} + \underline{I}^T \underline{\underline{C_2}} \underline{I} = [c_1 \quad \cdots \quad c_n] \begin{bmatrix} I_1 \\ \vdots \\ I_n \end{bmatrix} + [I_1 \quad \cdots \quad I_n] \begin{bmatrix} c_{1,1} & \cdots & c_{1,n} \\ \vdots & \ddots & \vdots \\ c_{n,1} & \cdots & c_{n,n} \end{bmatrix} \begin{bmatrix} I_1 \\ \vdots \\ I_n \end{bmatrix}, \tag{13}$$

where $I_i$ is a short form of $I_{common,i}$, $\underline{c_1}$ is the vector listing all linear coefficients, and $\underline{\underline{C_2}}$ is a matrix containing all second-order coefficients. Higher-order multivariate polynomials can either be expressed succinctly in terms of higher-order tensors or monomial by monomial. Table 1 provides a few example material systems and their form of the coefficients.

Table 1 Examples of beam coupling configurations under the polynomial photoexcitation model.

| Configuration description | Polynomial form | Linear coefficient vector $\underline{c}_1$ | Quadratic coefficient matrix $\underline{C}_2$ |
|---|---|---|---|
| A configuration with three mutually incoherent beams at the same wavelength exciting one photoinitiator system | $f = \alpha(I_1 + I_2 + I_3)$, where $\alpha$ is the combined coefficient that encapsulates the linear absorption coefficient and quantum yield of the photoinitiator at the wavelength of the light. | $\alpha \begin{bmatrix} 1 \\ 1 \\ 1 \end{bmatrix}$ | $\begin{bmatrix} 0 & 0 & 0 \\ 0 & 0 & 0 \\ 0 & 0 & 0 \end{bmatrix}$ |
| A two-beam photoinitiation and photoinhibition configuration[14–16], where beam 1 at wavelength 1 excites the photoinitiator and the beam 2 at wavelength 2 excites the photoinhibitor | $f = \alpha_1 I_1 - \alpha_2 I_2$, where $\alpha_1$ is the combined coefficient (encapsulating linear absorption coefficient and quantum yield) of the photoinitiator at wavelength 1, and $\alpha_2$ is the combined coefficient (encapsulating linear absorption coefficient, quantum yield, and inhibition reaction rate constants) of the photoinhibitor at wavelength 2. | $\begin{bmatrix} \alpha_1 \\ -\alpha_2 \end{bmatrix}$ | $\begin{bmatrix} 0 & 0 & 0 \\ 0 & 0 & 0 \\ 0 & 0 & 0 \end{bmatrix}$ |
| A two-beam two-photon lithography configuration where two beams at different wavelengths excite the photoinitiator with degenerate and non-degenerate two-photon absorption | $f = \alpha_{DTPA,1} I_1^2 + \alpha_{DTPA,2} I_2^2 + 2\alpha_{NDTPA} I_1 I_2$, where $\alpha_{DTPA,1}$ and $\alpha_{DTPA,2}$ are the combined coefficients that encapsulate the degenerate two-photon absorption cross-section and quantum yield at wavelength 1 and 2, respectively, and $\alpha_{NDTPA}$ is the combined coefficient that encapsulates the non-degenerate two-photon absorption cross-section and quantum yield. | $\begin{bmatrix} 0 \\ 0 \end{bmatrix}$ | $\begin{bmatrix} \alpha_{DTPA,1} & \alpha_{NDTPA} \\ \alpha_{NDTPA} & \alpha_{DTPA,2} \end{bmatrix}$ |

Note that in the second example of Table 1, the excited photoinitiator and excited photoinhibitor are coupled by chemical reaction. In the current framework, the coupling between the excited chemical species is also included in the "effective photoexcitation" model $\mathfrak{P}$, despite the naming of the latter. For simplicity, the material response model $\mathcal{M}$ (as a univariate function of photoexcitation $f$) only models the effect of net photoexcitation. An alternate and more sophisticated framework can be constructed to model the material response as a function of multiple types of photoexcitation.

## 3    OPTIMIZATION DEMONSTRATIONS AND ANALYSES

To showcase the efficacy of this optimization framework, optimizations are performed with the three example configurations described in Table 1 and the results are visualized in the following subsections. In addition, subsection 3.4 demonstrates the optimization for an additional holo-tomographic configuration where the holographic and tomographic light crafting capability is combined.

In each of these cases, a joint optimization and a decoupled optimization are performed. The joint optimization predicts the coupled performance of all the beams and updates all variables in parallel in every optimization step. This approach leverages all the tooling provided in the proposed framework and permits better coordination between beams. The decoupled optimization takes an alternate approach where it partitions the overall optimization problem into sub-problems where only one beam is considered at a time. This approach minimizes memory requirement in computation, but it relies on an educated guess to manually construct separate optimization targets for each beam and precludes beam coordination.

The solutions obtained from these two approaches are compared to highlight the performance benefit of fully coupled optimization.

All demonstrations are optimized towards the same response target $f_T$. The response target is a binary indicator function of space, which has a unity degree of conversion inside the set and zero outside. Geometrically, the positive response target is an octet truss structure. The span of the common coordinate grid is set to be about 86 $\mu m$ in all dimensions such that it matches the lateral field-of-view of beam 1 in all examples. In the examples, beam 1 is always at 473nm, 0.7 NA, and phase modulated at 256 x 256 resolution and a pixel size of 32 $\mu m$. The response target has a physical size that fits in 75% of the common coordinate grid. Figure 5 (a)-(b) shows the geometry of the response target from an oblique 3D view and on the z mid-plane. Figure 5 (c)-(d) graphically illustrates the beams' orientation relative to the target in each demonstration case. For simplicity and intuitive interpretation of the results, all optimizations assume an identity material response ($f_m = f$).

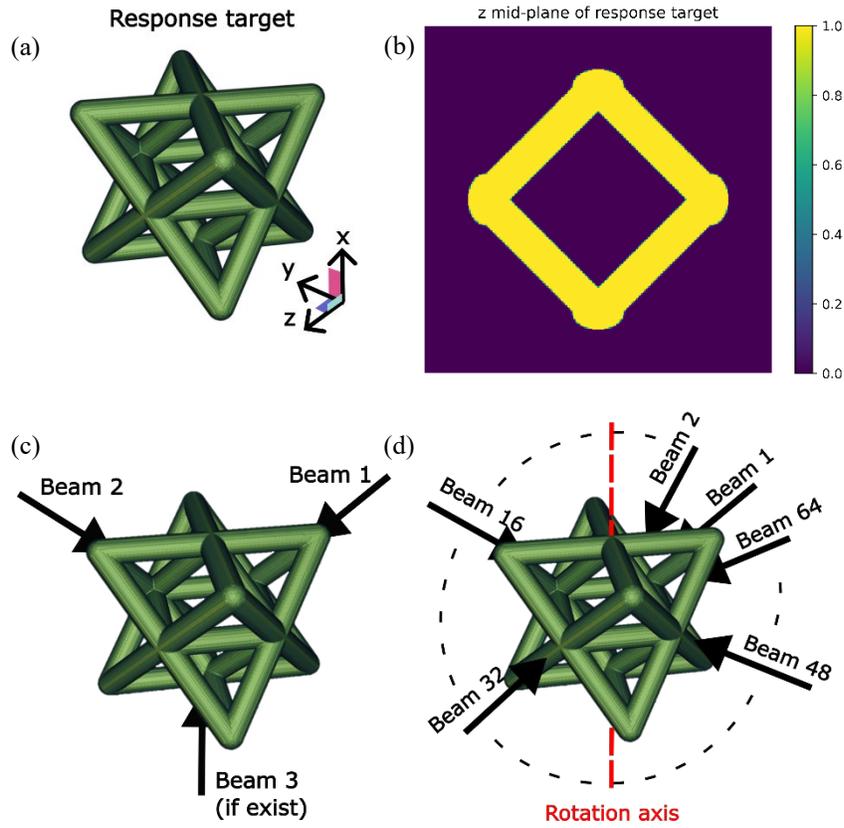

Figure 5 (a) Binary response target and coordinate axes, (b) z mid-plane (x-y cross-section) of response target, (c) three-beam and two-beam arrangement for demonstration 1 to 3, and (d) time-multiplexed beam arrangement for holo-tomographic patterning in demonstration 4.

Ideally, the first iterate of the solution can be constructed by attempting to solve the equation $f_m(\mathbf{r}) = f_T(\mathbf{r})$ which would minimize the loss function. However, the forward model of this optimization problem is not invertible, and exact solutions may not exist. Therefore, the demonstration in this work constructs the initial solution heuristically by applying known information about the system.

The initialization process first determines an ideal intensity field $I_{ideal,common,i}$ for each beam in the common coordinate grid such that their collective response would approximate $f_T$. The following subsections describe the case-specific method of constructing the ideal intensity field $I_{ideal,common,i}$. Then, the initialization process interpolates each ideal intensity field onto the respective beam coordinate grids to obtain a resampled intensity $I_{ideal,beam,i}$. Next, it constructs an artificial light field $u_{a,h,i}$ with a squared amplitude equal to $I_{ideal,beam,i}$ and a spatially variant random phase in the beam coordinate grid. Finally, the light fields $u_{a,h,i}$ at each $z_i$-plane are propagated back to the phase modulation plane as $u_{a,pm,i}$ using the following operation:

$$u_{a,pm,i}(x',y',z_i) = P_{c,i}^{-1} u_{a,h,i}(x_i,y_i,z_i), \tag{14}$$

where $P_{c,i}^{-1} = H^{-1}\mathcal{F}^{-1}$. The amplitude of the light field $|u_{a,pm,i}(x',y',z_i)|$ is averaged over $x',y',z_i$ and serves as the initial iterate of $\bar{a}_i$. The phase of the $z_i$-averaged light field $\angle \bar{u}_{a,pm,i}(x',y')$ serves as the initial iterate of $\phi_i(x',y')$. Explicitly, the $z_i$-averaged light field $\bar{u}_{a,pm,i}(x',y')$ is computed as:

$$\bar{u}_{a,pm,i}(x',y') = \frac{1}{n_{z_i}} \sum_{z_i} u_{a,pm,i}(x',y',z_i), \tag{15}$$

where $n_{z_i}$ is the number of $z_i$ planes present in the beam's discrete coordinate grid and the summation is performed over all $z_i$ planes.

In the propagation step of $P_{c,i}^{-1}$, the results of $H^{-1}$ is programmatically set to zero at far $x',y'$ locations where $H$ is zero since these regions correspond to non-propagating (evanescent) waves in ASM as mentioned in section 2.2. It should also be noted that the above interpolation operation (which resample $I_{ideal,common,i}$ from the common coordinate grid to the individual beam's coordinate grid) is neither inverse nor adjoint of $N_i$ in eq. (11) which resamples quantities in the opposite direction.

All plots in the following subsections use $f_m = 0$ and $f_m = 1$ as the bounds of color mapping and opacity mapping. Values outside of this bound are visualized as the corresponding boundary values. The colormap is perceptually uniform and the local opacity is directly proportional to response values.

### 3.1 Three-beam additive configuration

In this configuration, three beams at wavelengths 473nm are projecting light to the material in a mutually orthogonal arrangement. This beam arrangement is similar to that in single-shot VAM using orthogonal projections[39] where the beams additively compose the total reconstruction and each beam encodes a different set of spatial frequencies. Comparatively, this demonstration uses a much higher NA of 0.7 and a depth of focus that is much smaller than the target volume. The high NA and short depth of focus allow each of the holographic beams to focus on a range of axial positions almost independently.

This configuration uses the first example photoexcitation model in Table 1 with $\alpha = 1$. As mentioned, material response in the examples has an identity relationship with photoexcitation. During the initialization, the ideal intensity field is simply set as $I_{ideal,common,i} = f_T/3\alpha = f_T/3$ for $i = \{1,2,3\}$. In other words, the ideal intensity fields of each beam are assumed to contribute to the total required photoexcitation equally.

The upper and lower subfigure on the left column of Figure 6 shows the 3D volumetric plot and 2D midplane slice plot of the material response $f_m$ obtained from the initial guess solution, respectively. The right column of Figure 6 plots the material response obtained from multi-beam joint optimization. A decoupled optimization is performed where all the beams are individually optimized in their respective one-beam system without any information about other beams' contributions. The response targets for the individual beams are set as one-third of that in joint optimization, i.e. $f_{T,decoupled} = f_T/3$. After optimization, the intensities are combined through the actual photoexcitation model to obtain the combined response $f_{m,combined} = I_{decoupled,1} + I_{decoupled,2} + I_{decoupled,3}$. The combined response is visualized on the two subfigures in the middle columns of Figure 6. Both the joint optimization and decoupled optimization are run for 5000 iterations.

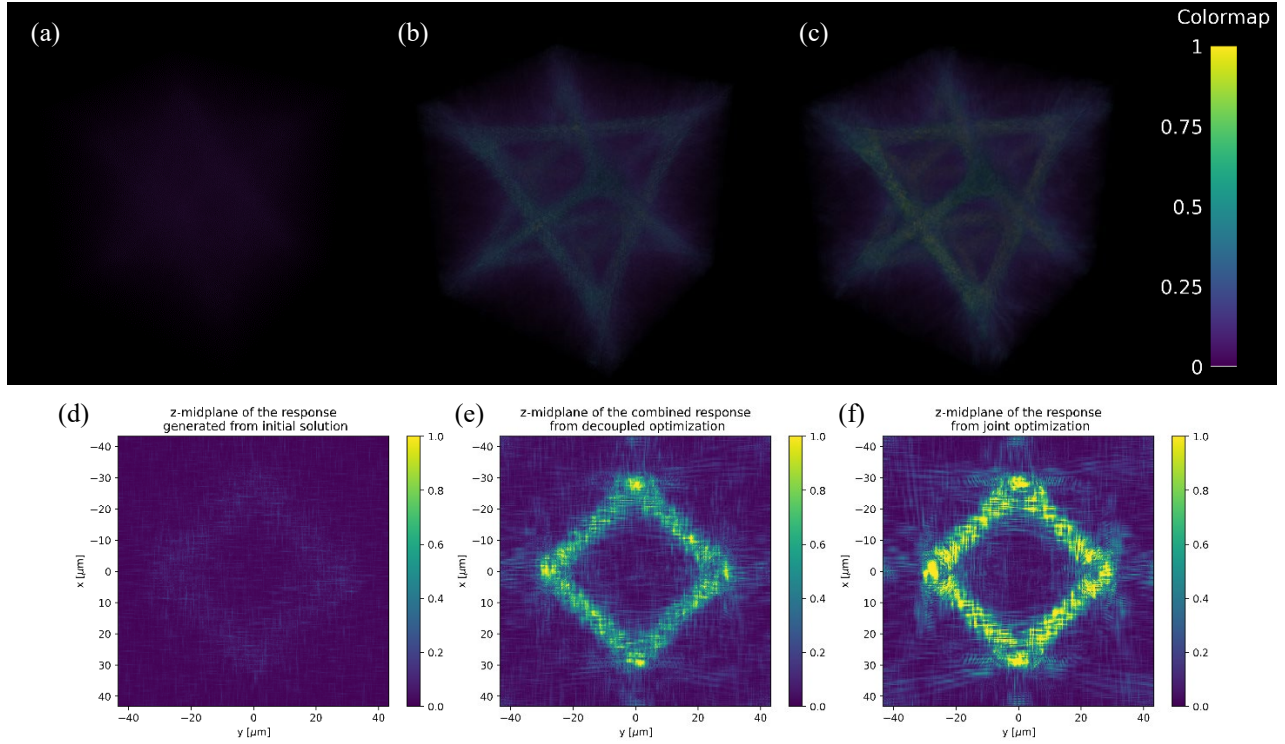

Figure 6 (a)-(c) 3D oblique view of the response generated by (a) initial solution in joint optimization, (b) final solution in decoupled optimization, and (3) final solution in joint optimization, respectively. (d)-(f) z-midplane cross-section of response generated by (d) initial solution in joint optimization, (e) final solution in decoupled optimization, and (f) final solution in joint optimization, respectively.

Both qualitatively and quantitively, the joint optimization performed better than the decoupled optimization. The response contrast in the joint optimization is visibly higher. The combined solution in decoupled optimization has a loss value that is 14% higher than that of joint optimization. Both optimizations significantly increase the beams' amplitude from their initial guesses. In the decoupled optimization, the mean beam amplitudes for each beam increase similarly from roughly 0.13 to about 0.27 in 5000 iterations. In the joint optimization, the amplitudes for all three beams increase from 0.13 to 0.30 in 5000 iterations.

As mentioned in section 2.3, the simulation assumes there is no mutual coherence between the beams, despite the beams being set at the same wavelength. The results obtained here correspond to the situations where the beams are out of spatial or temporal coherence with each other or separated in time.

### 3.2   Two-beam subtractive configuration

The two-beam subtractive configuration demonstrates the antagonistic interaction between optical exposure of different wavelengths. Using the photoinitiation/photoinhibition material system[14–16], the first beam at 473nm wavelength provides photoexcitation to polymerize the material while the second beam at 360nm wavelength provides photoexcitation to inhibit polymerization. This example corresponds to the second row in Table 1 and uses a unity value for both linear polynomial coefficients. The effective photoexcitation takes the form of $f = I_1 - I_2$. The second beam is set to have a NA of 0.54 such that its field-of-view matches that of first beam, which has a longer wavelength and a NA of 0.7.

In both joint optimization and decoupled optimization, the initialization process uses $I_{ideal,common,1} = f_T$ for the first beam and $I_{ideal,common,2} = f_T'$ for the second beam, where $f_T'(\underline{r}) = \max_{\underline{r_0}} f_T(\underline{r_0}) - f_T(\underline{r})$ is constructed as a positive real-valued complement of $f_T$. The joint optimization directly optimizes the resultant response towards $f_T$ since the subtraction relationship of the two beams is taken into account by the photoexcitation model. The decoupled optimization optimizes two one-beam systems separately and assigns $f_T$ and $f_T'$ as the response target $f_{T,decoupled,i}$ for beam 1 and beam 2 respectively. The subtraction of the beam intensities is only performed after the optimization to compute the combined response $f_{m,combined} = I_{decoupled,1} - I_{decoupled,2}$. Figure 7 plots the responses from the two optimizations.

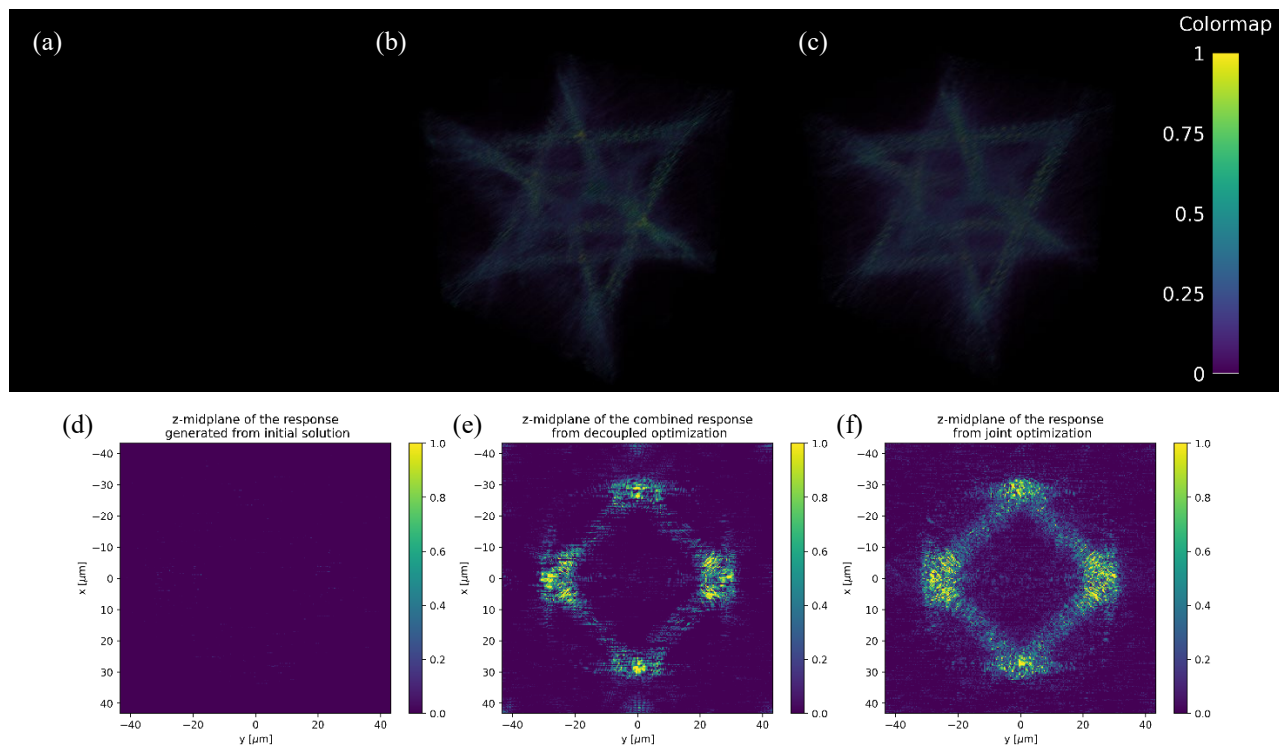

Figure 7 (a)-(c) 3D oblique view of the response generated by (a) initial solution in joint optimization, (b) final solution in decoupled optimization, and (3) final solution in joint optimization, respectively. (d)-(f) z-midplane cross-section of response generated by (d) initial solution in joint optimization, (e) final solution in decoupled optimization, and (f) final solution in joint optimization, respectively.

In this example, the joint optimization performed significantly better than the decoupled optimization. The jointly optimized response reproduced the struts of the target much more reliably. The combined response from decoupled optimization has a loss value that is 47% higher than that of joint optimization. Both optimizations start with a solution that has a much stronger inhibition beam (with a mean amplitude of 0.99) than the initiation beam (with a mean amplitude of 0.23) and therefore produce almost no positive response initially. To steer the response closer to the target, both optimizations increase the initiation beam amplitude to about 0.46. However, the joint optimization results in a lower final inhibition beam amplitude (at 0.34) relative to the decoupled optimization (at 0.42). Therefore, the solution from joint optimization is not only more accurate but also more energy efficient. Since energy efficiency is not explicitly formulated into the loss function, this is likely a side benefit of the strategic placement of antagonistic excitations.

### 3.3 Two-beam two-photon configuration

The two-beam two-photon configuration uses two beams of different wavelengths to trigger two-photon absorption (TPA) of the photoinitiator. This configuration corresponds to the last row of Table 1. The wavelengths of beam 1 and beam 2 are set as 473nm and 532nm respectively to keep the patterning volume close to other examples.

In this two-beam setup, there are two possible modes of TPA, namely the degenerate TPA (DTPA) and non-degenerate TPA (NDTPA)[40,41]. DTPA occurs when the molecule of concern absorbs two photons of the same wavelength simultaneously (during the lifetime of the virtual intermediate state). With the above choice of wavelengths, it corresponds to the simultaneous absorption of two 473nm photons or two 532nm photons. In comparison, NDTPA occurs when the molecule absorbs one 473nm and one 532nm photon simultaneously. The polynomial coupling coefficients of these TPA may vary widely and depend on the choice of molecule and excitation wavelengths. In this qualitative demonstration, all polynomial coefficients are assumed to be 1 and this setting represents a strongly coupled excitation system where $f = I_1^2 + I_2^2 + 2I_1I_2 = (I_1 + I_2)^2$. The initialization process set $I_{ideal,common,i}$ to equal $\sqrt{f_T}/2$ for both beams in joint optimization.

The decoupled optimization optimizes the two beams individually as two one-beam systems. Within each one-beam system, the photoexcitation has a quadratic dependence on intensity ($f_{decoupled} = I_{decoupled}^2$). The response target of the one-beam systems is set as $f_{T,decoupled} = f_T/4$ such that their optimized intensities (which are ideally close to $\sqrt{f_T}/2$) would combine in the actual photoexcitation model $f = (I_1 + I_2)^2$ to make up $f_T$ after optimization. The initialization variable $I_{ideal,common,i}$ is set to be the square root of the response target ($\sqrt{f_T}/2$) and coincides that in the joint optimization. A combined response is pieced together after optimization with $f_{m,decoupled} = (I_{decoupled,1} + I_{decoupled,2})^2$. The responses from the two optimizations are plotted in Figure 8.

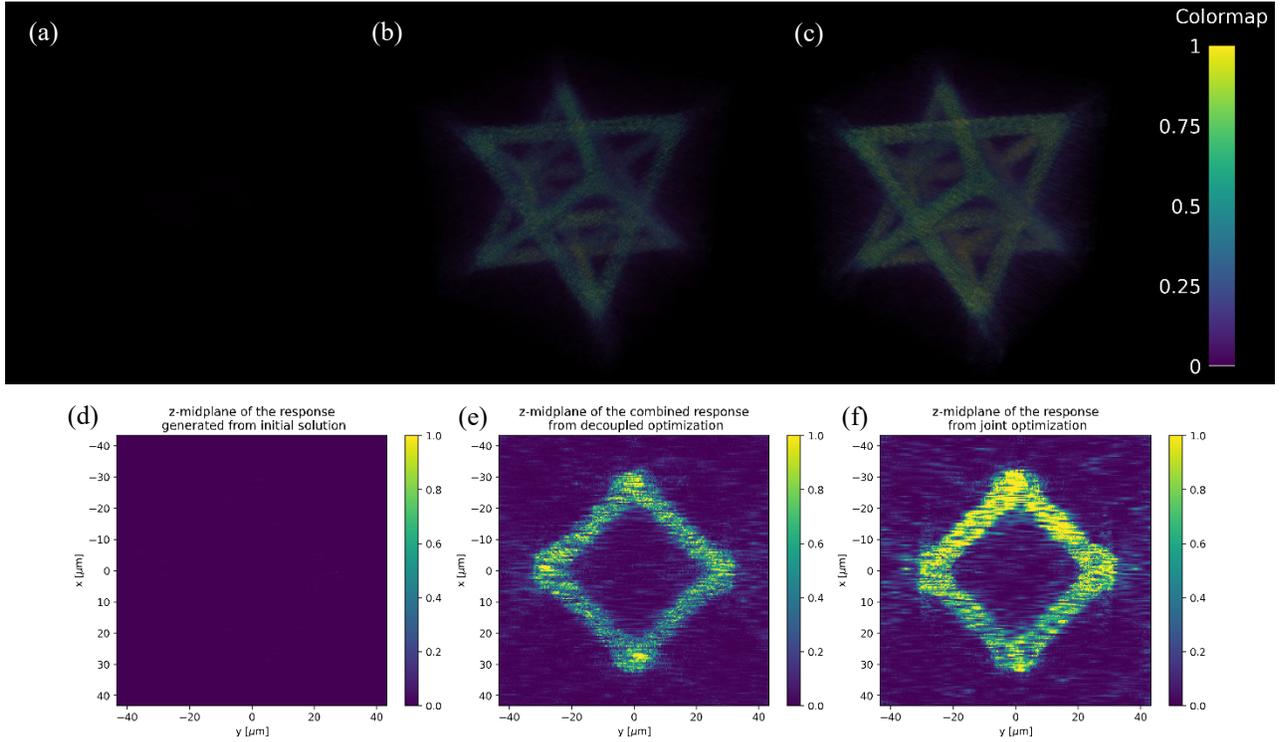

Figure 8 (a)-(c) 3D oblique view of the response generated by (a) initial solution in joint optimization, (b) final solution in decoupled optimization, and (3) final solution in joint optimization, respectively. (d)-(f) z-midplane cross-section of response generated by (d) initial solution in joint optimization, (e) final solution in decoupled optimization, and (f) final solution in joint optimization, respectively.

The response from the initial guess has a magnitude much lower than the target. This underestimation can be attributed to several factors. Firstly, the initial guess of the mean beam amplitude $\bar{a}_i$ may not conserve energy from the guess light field $u_{a,h,i}$ in eq. (14). Secondly, the initial guess hologram may be inefficient in localizing the energy over the positive target (where $f_T > 0$) and instead spreading out a sizable amount of energy into the background. Thirdly, the photoexcitation's quadratic dependence on intensity further exacerbates the underestimation since a squaring operation pushes positive non-unity values further away from unity. These compounding effects make the response effectively invisible in Figure 8 (a) and (d).

Despite the initial underestimation, both optimizations successfully generate a solution that is close to the target in magnitude. Comparing the two response profiles, the jointly optimized response has a higher fidelity. The loss value of the combined response from decoupled optimization is 33% higher than that from joint optimization.

## 3.4 Time-multiplexed holo-tomographic configuration

Existing demonstrations of tomographic VAM project planar images and do not have the point-by-point axial focusing capability offered by holographic reconstruction. The holo-tomographic patterning configuration in this example represents a convergence of holographic and tomographic VAM where the setup leverages both axial focusing and relative rotation to localize exposure. The setup physically contains one beam that is rotating relative to the material. The beam delivers a different holographic projection at different times and accumulates exposure. In this optimization framework, such a time-multiplexed beam is decomposed into many independently modulated beam instances.

This example uses $N = 64$ beams spanning a full 360° rotation and the beams deliver single-photon photoexcitation additively. In the photoexcitation model, the linear coupling coefficient is simply a vector with all elements being one:

$$\underline{c_1} = [1, \dots, 1]^T. \tag{16}$$

To simulate a scanning setup with a constant-power light source, the mean beam amplitudes $\bar{a}_i$ of all the beams are treated collectively as one optimization variable. In the initialization of both joint optimization and decoupled optimization, $I_{ideal,common,i}$ is evaluated as $\frac{f_T}{N} = \frac{f_T}{64}$. The response obtained by jointly optimizing the 64 beams is plotted in the last column of Figure 9.

The decoupled optimization uses $f_{T,decoupled} = f_T/64$ as the response target for each one-beam optimization. The naïve approach of performing the same number of iterations (5000) for 64 beams individually would require extensive computational time. By taking advantage of the two-fold rotational symmetry of the response target about the chosen rotation axis, only the first 32 beams spanning $[0°, 180°)$ angular range are selected as reference and are individually optimized. In the computation of the combined response, the other half of the beams repeat the phase modulation values obtained from the 32 beams in sequential order. A total of 64 beams are propagated to the volume and their common beam amplitude is taken as the average of the 32 individually optimized mean beam amplitudes. The combined response of individually optimizing the component beams is plotted in the middle column of Figure 9.

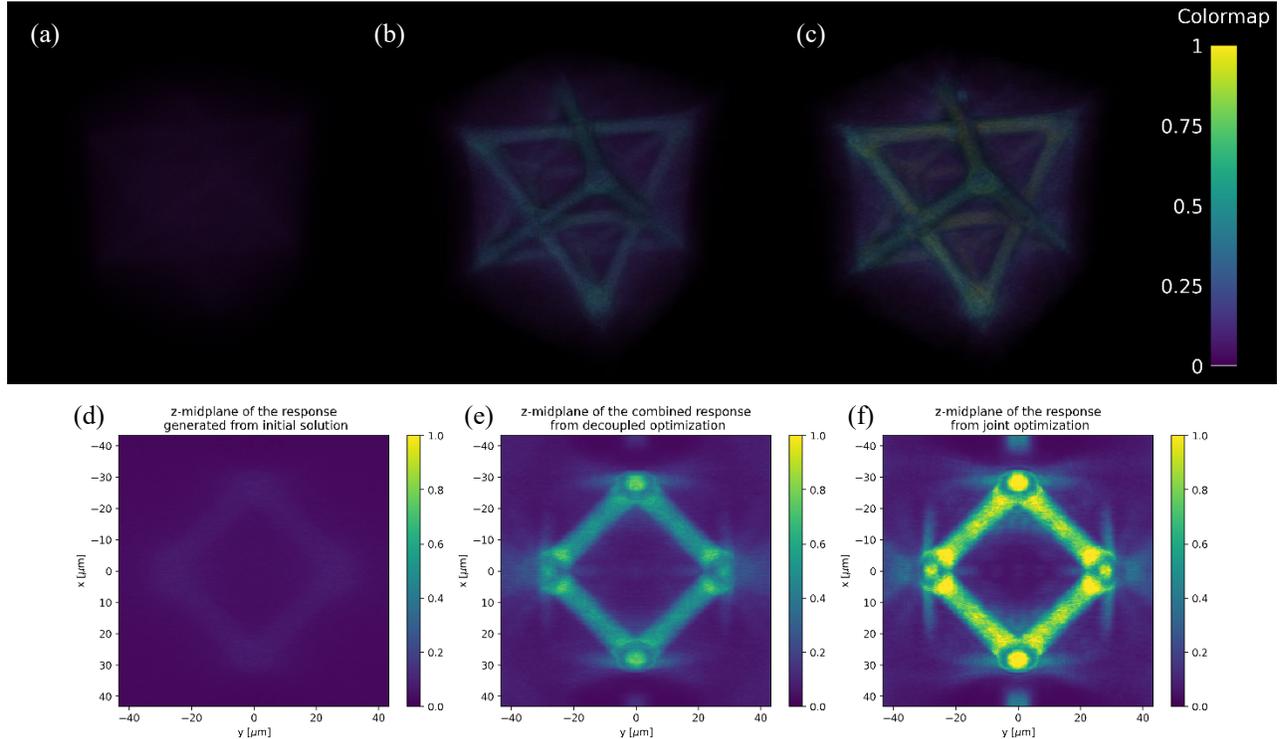

Figure 9 (a)-(c) 3D oblique view of the response generated by (a) initial solution in joint optimization, (b) final solution in decoupled optimization, and (3) final solution in joint optimization, respectively. (d)-(f) z-midplane cross-section of response generated by (d) initial solution in joint optimization, (e) final solution in decoupled optimization, and (f) final solution in joint optimization, respectively.

Among other demonstrations, the holo-tomographic configuration achieved the highest reconstruction fidelity in the optimized solutions. This can be concluded both from visual inspection and from its lowest loss value. The reconstructed responses in this configuration are substantially more uniform due to the superposition of a high number of beams that smooth out the speckles. From Figure 9 (e) and (f), the background exposure from adjacent features is clearly visible near the four nodes. The two nodes that are at $y = 0$ has a different exposure pattern than the other two at $x = 0$. As the target has four-fold rotational symmetry in this plane, this difference in pattern is likely driven by the choice of the rotation axis ($x$).

Similar to other examples, joint optimization yields a solution that is superior to that generated by decoupled optimization. The loss value of the response in the latter is 36% higher than that in the former. Visual inspection suggests that the higher loss of the combined solution can partially be attributed to the overall magnitude mismatch. This magnitude mismatch indicates that individual one-beam optimizations tend to converge to a sub-optimal mean beam amplitude $\bar{a}_i$ since it has no information about the overall response and the overall response target. In this simple case, the solution in the decoupled optimization can potentially be improved post-optimization with a magnitude scaling. However, it should be noted that this simple scaling is not viable when the photoexcitation model or material response model is non-linear. In general, a systematic search of the optimal beam power is still needed during the iterative optimization process. This is one of the key reasons underlying the consistently superior performance offered by the joint optimization approach.

# 4    CONCLUSION

An optimization framework is proposed in this work to simultaneously optimize both the phase and amplitude variables of multiple holographic projections for high-fidelity volumetric printing. The framework addressed multiple modeling and computational needs that traditional phase retrieval methods fail to meet in the context of multi-beam holographic VAM. At a high level, this framework uses a general loss function to specify accuracy requirements and emphases flexibly. Components of the framework are designed to lower computational costs and keep the high-dimensional problem tractable. The framework uses a coherent propagation method in Fourier optics to compute the holographic light field efficiently. A resampling step in the framework allows the individual holograms and target quantities to be evaluated at their native sampling rates. The use of automatically differentiable operations facilitates the evaluation of loss gradient and performant gradient-based optimization on GPU. The intuitive low-order polynomial photoexcitation model in the framework is capable of capturing a wide variety of non-linear exposure coupling effects in multi-photon and multi-color material systems. In all test cases, the superior reconstruction quality from joint optimization highlighted the importance of proper modeling of the coupled contributions. Among other configurations, the time-multiplexed holo-tomographic configuration delivers the best reconstruction fidelity and represents a natural extension of tomographic VAM for high NA microscale fabrication. Promising directions for development include explorations of better initialization strategies, speckle reduction regularization methods, and application of adaptive stepping methods.